\documentclass[aps,pra,twocolumn,groupedaddress,showpacs]{revtex4}
\usepackage{amssymb}
\usepackage{amsmath}
\usepackage{graphicx}

\begin{document}

\title{Quantum communication on closed time-like curves: a protocol for exchanging, protecting and storing qubit states}

\author{Ian T. Durham}
\email[]{idurham@anselm.edu}
\affiliation{Department of Physics, Saint Anselm College, Manchester, NH 03102}
\date{\today}

\begin{abstract}
We describe a weaker consistency condition for qubits on closed time-like curves (CTCs) and define a new quantity we call a \emph{ctcbit} that provides a means for quantifying a qubit on CTC as a shared resource.  We describe a simple protocol for the sharing of information that is similar to quantum teleportation but does not require an entangled particle pair or \emph{ebit}.  The nature of CTCs also serves as a way to protect a qubit state.  While there is the appearance that the given resource is free, we employ a non-Hausdorff topology to prevent any limitless information exchanges.  While the reality of CTCs is highly speculative, the present paper provides a manner by which quantum informational methods may be employed to study such problems and may ultimately prove useful in studying quantum gravity.
\end{abstract}


\maketitle

\section{PRELIMINARIES}
Quantum computational methods have been proposed as a way to solve a number of problems generally thought to be either intractable with classical computational methods.  One set of such problems includes problems in quantum gravity.  Among these are the quantum behavior of closed time-like curves (CTCs).  CTCs naturally arise from the construction of wormholes \cite{hawking92} and the latter were first seriously considered in the modern literature by Morris, Thorne, and Yurtsever \cite{thorne88}.  Deutsch first considered the merger of quantum computational methods and CTCs demonstrating that quantum computation in the presence of CTCs always allows self-consistent evolution \cite{deutsch91}.  Classically similar ideas were later discussed by Brun \cite{brun02} before Bacon put them on a slightly firmer ground by coupling the CTC qubits to chronology-respecting qubits \cite{bacon04}.  Ralph \cite{ralph07} recently proposed an alternative to BaconÕs treatment.

While these analyses are theoretically interesting, there is no conclusive evidence that CTCs physically exist in nature \cite{hawking92,kim91,visser,deser} though it is entirely possible that we will come closer to a determination of this as new experiments and theoretical frameworks are suggested  \cite{davies07,simeone}.  Nonetheless, the theoretical work continues unabated.  A particularly daunting problem is that of the initial value problem on spacetimes with CTCs \cite{deutsch91,friedman90,friedman91,hartle94,goldwirth,politzer94a,politzer94b,cassidy,hawking95}.  One of BaconÕs goals was to tackle this very problem while simultaneously using quantum evolution in the presence of CTCs to efficiently solve NP-complete problems.

At the same time, we have come to understand information on the basis of shared resources such as \emph{cbits}, \emph{ebits}, \emph{cobits}, \emph{refbits}, and, of course, \emph{qubits} \cite{vanenk,harrow,nc}.  Given BaconÕs results, it seems logical to ask whether this coupling of qubits on CTCs with chronology-respecting qubits can be exploited in some manner for the purposes of exchanging information; in short, can a shared resource be developed from this?

The use of CTCs, as it turns out, provides us with some advantages over typical protocols such as the quantum teleportation protocol \cite{bb93,bez,riebe,barrett} that has been described by Loepp and Wootters as a sort of Òdestructive faxingÓ in which an unknown or general quantum state can be transmitted over long distances without any loss of information \cite{loepp}.  Quantum teleportation, however, requires the use of an entangled pair that we refer to as an \emph{ebit} (defined below). Protocols on CTCs can accomplish a very similar end result but without \emph{ebits}, though there are fundamental limitations in certain instances.  In addition, in order to avoid any possibility of a limitless and freely available resource we introduce a non-Hausdorff topology.

\section{NOTATION AND BASIC DEFINITIONS}
We follow the same basic notation as Ref. \cite{vanenk} in which the terms qubit and \emph{qubit} have slightly different meanings. We define a qubit as being a physical entity with some binary property.  That is this property can be represented by, at most two, orthogonal pure states or by mixed states that are always some superposition of these two orthogonal pure states.  Examples of qubits, of course, include two-level atoms or molecules as well as polarized photons.  We then define a \emph{qubit} as being a communication resource that is equivalent to sending a physical qubit over a noiseless channel.  Again this is the definition given in Ref. \cite{vanenk}.  Thus we note that the italicized form represents the shared resource.  To clarify this notation, let us define an \emph{ebit} as being the resource of Alice and Bob sharing a maximally entangled state of a particular form for use in quantum communication.  An \emph{ebit} is considered to be a unit of entanglement, i.e. it provides a manner by which entanglement may be measured.  Note that in this article we do not distinguish between capitalized and non-capitalized forms of these resources, unlike Ref. \cite{vanenk}, to which we refer the reader for a fuller discussion.  We also note that, despite MerminÕs valid orthographical point about qubits (i.e. that they should be qbits) \cite{mermin}, we will maintain the more widely accepted spelling.

\section{CTC EVOLUTION AND STRUCTURE}
Wormholes (and the associated CTCs) do not quantum mechanically evolve via the Schršdinger equation.  In fact the CTCs themselves do not evolve at all temporally since they obey what is known as the Wheeler-DeWitt equation (see Ref. \cite{visser}), \begin{equation}H|\psi\rangle=0\end{equation}where \emph{H} is the Hamiltonian operator.  The only exact solution that is known for the Wheeler-DeWitt equation is one in which there is no mass in the wormhole throat, though the WKB approximation can be used to give approximate solutions in cases where mass is present \cite{visser}.  This ultimately leads one to consider consistency conditions, i.e. what does this equation mean in terms of the evolution of the CTC qubit.  Note that the Hilbert space, \begin{math}\mathcal{H}\end{math} of the combined system is the tensor product of the individual Hilbert spaces, \begin{math}\mathcal{H}_1\otimes\mathcal{H}_2\end{math}.
\\\\
\noindent\textbf{Strong consistency condition}  The strongest consistency condition, then, does not allow the CTC qubit's wavefunction to evolve under unitary transformations.  In other words, \begin{equation}
(|s^{\prime}\rangle \otimes |ctc\rangle)=U(|s\rangle \otimes |ctc\rangle)
\end{equation} where we notice that the state of the CTC qubit, \begin{math}|ctc\rangle\end{math}, does not change but the state of the chronology respecting qubit does, evolving from \begin{math}|s\rangle\end{math} to \begin{math}|s^{\prime}\rangle\end{math}. A little algebra will show that this consistency condition requires that the chronology-respecting qubit and the CTC qubit must be in the same basis.
\\\\
\noindent\textbf{Deutsch-Bacon consistency condition}  A slightly weaker consistency condition was proposed by Deutsch \cite{deutsch91} and further studied by Bacon \cite{bacon04}.  It merely requires the \emph{density operator} of the CTC qubit remain unchanged,\begin{equation}\rho=\textrm{Tr}_\textrm{{\scriptsize A}}\left[U\left(\rho_{in}\otimes\rho\right)U^\dag\right]\end{equation}
where a partial trace is taken over the chronology-respecting qubit.  This is a weaker condition because it is possible for more than one wavefunction to be represented by a single density operator \cite{nc}.  Note that this \emph{does not} imply that the CTC qubit and the chronology-respecting qubit are entangled.  In fact it is highly likely (though this has never been formally proven) that the equation \emph{prevents} any such entanglement from occurring.  Since CTCs are often associated with very strong gravitational fields, at least in portions of the wormhole throat, this would be consistent with the results found in \cite{fuentes05,alsingmilburn,haydenpreskill}.

\subsection{Interpreting the Deutsch-Bacon consistency condition}
It would seem as if equation (3) indicates that the CTC qubit's state \emph{depends on} the state of the chronology-respecting qubit and, if so, would present a paradox.  However, note that the action of the unitary in equation (3) is to couple the two qubits.  It is entirely within reason to, instead, interpret equation (3), then, as simply limiting what states represented by \begin{math}\rho_{in}\end{math} may be properly coupled to the CTC qubit.  In other words, instead of viewing \begin{math}\rho_{in}\end{math} as the \emph{independent} 'variable,' so to speak, with \begin{math}\rho\end{math} acting as the \emph{dependent} 'variable,' we instead do the opposite and assume \begin{math}\rho\end{math} to be independent while \begin{math}\rho_{in}\end{math} is dependent.  Attempting to couple incompatible states would simply in noise, i.e. meaningless output.

In analyzing equation (3) then from this viewpoint, we find that the CTC qubits may only be coupled to chronology-respecting qubits whose states are pure.  This is because only the pure states can be traced out of equation (3).  We also find that the choice of unitary \emph{further} limits what states will work.  Consider, for example, the action of a controlled-rotation gate whose unitary is \begin{math}\textrm{\textbf{U}}=|00\rangle\langle00|+|01\rangle\langle01|+|10\rangle\langle10|+i|11\rangle\langle11|\end{math}.  We define \begin{math}\rho \equiv |e_{j}\rangle\langle e_{k}|\end{math} and note that, assuming the chronology-respecting qubit's state is pure, a great deal of tedious algebra reduces equation (3) to 
\begin{eqnarray*}
 |e_{j}\rangle\langle e_{k}| & = a|0\rangle\langle0|e_{j}\rangle\langle e_{k}|0\rangle\langle0|+b|0\rangle\langle0|e_{j}\rangle\langle e_{k}|1\rangle\langle1| \\
& +c|1\rangle\langle1|e_{j}\rangle\langle e_{k}|0\rangle\langle0|+d|1\rangle\langle1|e_{j}\rangle\langle e_{k}|1\rangle\langle1| \\
& +ie|0\rangle\langle0|e_{j}\rangle\langle e_{k}|1\rangle\langle1|-ie|1\rangle\langle1|e_{j}\rangle\langle e_{k}|0\rangle\langle0|.
\end{eqnarray*}
Notice that this only holds if \begin{math}j=k\end{math} which implies the CTC qubit is also in a pure state.  As another example, consider the simple swap gate whose unitary is \begin{math}\textrm{\textbf{U}} = 00\rangle\langle00|+|01\rangle\langle10|+|10\rangle\langle01|+|11\rangle\langle11|\end{math}.  This ultimately implies, after much algebra, that
\begin{eqnarray*}
 |e_{j}\rangle\langle e_{k}| & = a|0\rangle\langle0|e_{j}\rangle\langle e_{k}|0\rangle\langle0|+b|1\rangle\langle1|e_{j}\rangle\langle e_{k}|1\rangle\langle1|.
\end{eqnarray*}
It is immediately obvious that the CTC qubit's state must be pure.

\section{WEAK CONSISTENCY CONDITION}
Technically, however, the Wheeler-DeWitt equation, from which equation (3) is essentially derived, only applies to the \emph{throat} of a wormhole.  In all practicality, the latter would only be a small portion of the entire CTC.  Thus a broader view of consistency really only requires that the CTC qubitÕs state is \emph{well-ordered and cyclic around a closed loop}.  That is, consider the CTC qubit in Figure 1.\begin{figure}
\begin{center}
\includegraphics{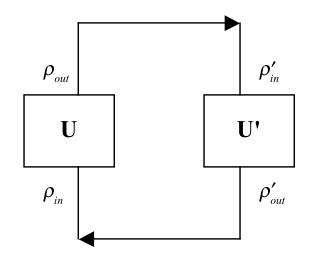}
\caption{This is a depiction of the CTC qubit state's evolution.  Consistency simply requires that it is always the same at a single point on the loop for different times and that it is well-ordered and cyclic.}
\label{ }
\end{center}
\end{figure}The CTC qubit's evolution is completely consistent as long as \begin{equation}\begin{array}{c}
      \rho_{out}=\rho_{in}^{\prime} \\
      \rho_{out}^{\prime}=\rho_{in}.
\end{array}\end{equation}

The weak consistency condition raises a number of questions.  We can imagine a host of potential problems relating to causality here if quantities, some of which are not even present in the above equations, could get altered.  This is perhaps why Bacon used the stronger requirement that the CTC qubit's density operator never change.  It avoided any such problems.  Nonetheless, there are some potential ways around some of these problems.  These problems are best understood by looking at an example.

Suppose we have, then, two parties, Alice and Bob, who share a CTC qubit.  Alice additionally has a chronology-respecting qubit in an unknown state and wishes to couple these two qubits via a unitary transformation of some sort as a means of communicating to Bob the state of this chronology-respecting qubit.  Alice needs to make some sort of measurement on the chronology-respecting qubit, preferably after the unitary transformation, before sending the results on to Bob.  Bob then uses these results to extract the information from the CTC qubit, perhaps by preparing an ancilla and coupling this to the CTC qubit, then measuring the ancilla after the unitary transformation (we discuss specifics below).

\subsection{Requirements of the weak consistency condition}
\noindent\textbf{Bob does not perform his transformation and measurement}  What happens if Bob fails to pass his qubits through the swap gate thus leaving the CTC qubit in some mixed state?  The states clearly would not exhibit behavior indicative of a closed loop.  Notice that throughout our protocol, measurements are only performed on the chronology-respecting qubit.  No measurements are performed on the CTC qubit itself.  Doing so would collapse the state.  The unitary transformation acts a bit like a measurement in that it changes the states of the qubits involves (i.e. it swaps them).  By not making one of these transformations we essentially leave the state 'collapsed' in one form.  In this instance, we could simply assume that the CTC itself collapses since it is intrinsically linked via the Wheeler-DeWitt equation to the qubit that traverses it.  This might also happen if either Alice or Bob deviated from the assigned spacetime coordinates, though it might depend perhaps partly on the location of the wormhole throat as well as other factors.
\\\\
\noindent\textbf{Could Bob learn the result of his measurement before making it?}  Since the curve is closed and time-like one might expect that Bob could simply send a message back in time to Alice informing her of the outcome of his measurement (recall that the initial chronology-respecting qubit state is unknown) so that she might, in turn, inform (or 'pre-inform') him so he never needs to make the measurement in the first place.  This of course would be paradoxical.  However, there are two assumptions we can make that prevent this from happening.  First we can assume that only the one CTC qubit is allowed to traverse the CTC.  This means Bob would have to encode the result of his measurement in the CTC qubit.  Doing so, however, would alter the CTC qubit's state such that it was not in its original state when Alice fed it into the initial swap gate.  Therefore, consistency \emph{prevents} Bob from surreptitiously informing himself of his own measurement result before that measurement has been made.
\\\\
\noindent\textbf{Epistemic knowledge on closed time-like curves}  In the above examples there is a very subtle assumption that is made about the flow of time.  Classically, information in time is assumed to flow linearly in one direction - in other words information can only be passed \emph{forward} in time.  CTCs appear to allow information to be passed \emph{backward} in time.  But on a CTC there are \emph{two} ways something might be considered to move backward in time.  If we visualize the CTC literally as a loop holding an object that is moving along this loop and we place this loop on a 'background' of linear time, as the object continues to move along the loop in the same direction (from the standpoint of the loop), it appears to reverse direction in reference to the background linear time.  On the other hand, the object might move backward relative to this background linear time by reversing it's motion on the loop itself.  It then is also moving backward in time according to the loop itself.  The latter is \emph{internally} inconsistent.  Thus we define an inherent chirality to temporal 'motion' on a CTC (see Figure 2). \begin{figure}
\begin{center}
\includegraphics{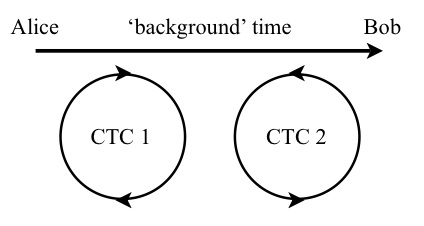}
\caption{Time moves from left to right for Alice and Bob.  For an object on a CTC to move backward in time relative to Alice and Bob it may either move as demonstrated by CTC 1 or by CTC 2.  For consistency, however, whenever the two curves come into contact they must be moving in the same direction, i.e. it wouldn't make sense to observe something moving backward in time while you yourself were moving forward in time.  If we assume the contact between the CTC and the linear time occurs at the top, this rules on CTC 2. }
\label{}
\end{center}
\end{figure}  If we assume the CTC time and the linear time must be moving in the same direction when they are in contact with one another (in the swap gate inside the laboratory) and we assume Bob's actions \emph{always} take place in Alice's future lightcone in their linear time, consistency is preserved.  In other words, Bob cannot utilize the CTC to send himself a signal from the past.  If he does the CTC must, by definition, collapse since consistency would not be preserved.

What this means is that, while an \emph{object} such as a qubit may seemingly move backward in time relative to our (or Alice's or Bob's) viewpoint, no \emph{knowledge} (i.e. information) may move backward in time.  An attempt to do so requires altering the CTC qubit's state again thus producing an inconsistent result (and thus collapsing the CTC).  Note that we assume here that classical information cannot be sent along the CTC from Bob to Alice.  This is a boundary condition we choose to make but is not necessarily a known requirement.  Note also that our above argument is entirely heuristic and solely based on the requirement of logical consistency and causality.  A more formal proof would be stronger and should be investigated.

There is another problem that we have glossed over until now.  That is, how does Alice learn what the state of the CTC qubit is?  She can't measure it without disturbing it and that would lead to an inconsistency.  One possible solution to this is for there to actually be a \emph{beam} of CTC qubits (e.g. photons, etc.) in random bases.  A protocol could then be carried out in much the same way as the BB84 protocol \cite{loepp}.  The measurement Alice makes after her unitary transformation is random.  With enough CTC qubits in the 'beam' we can expect her to guess correctly about half the time.  The only question is whether her incorrect guesses end up collapsing the CTC (see discussion of non-Hausdorff topologies below).
\\\\
\noindent\textbf{Can the CTC qubit be used more than once?}  This is an intriguing question since, once Alice and Bob have utilized the \emph{ctcbit}, it comes back to Alice's location at her original time.  In other words, Alice might seemingly have an infinite number of these CTC qubits on hand, all being different representations of the same CTC qubit.  This, of course, makes no sense.  But how is this problem avoided?  In other words, how is it possible to avoid having Alice's actions repeat themselves \emph{ad infinitum}?  Or, another way of looking at it, how can we ensure that once Alice and Bob utilize the CTC qubit, it cannot be utilized again by anyone else?

One way to do this is to introduce what is called a non-Hausdorff topology onto the CTC itself.  As Visser has suggested, a full theory of CTCs might require employing a non-Hausdorff topology \cite{visser}.  By employing such a topology we can ensure that a single, consistent history exists for each instance of the transmission.  Such a topology would have to be at the core of any complete theory of a branching spacetime, though Visser has also made it clear that branching spacetimes are not required for the purposes of consistency \cite{visser}.  Thus, while branching spacetimes are attractive with respect to this particular problem, they may not necessarily be required.  (Note that our employment of branching spacetimes in this context be construed as an endorsement of any particular interpretation of quantum mechanics.)  In order to explain a non-Hausdorff topology it is actually simpler to first define a Hausdorff topology.
\\\\
\noindent\textbf{Definition}  \emph{A topology}, \begin{math}\textrm{T}_2\end{math}, \emph{is Hausdorff if and only if for any two points} \begin{math}x_1\;and\;x_2,\;where\;x_1\neq x_2\end{math}, \emph{there exist open
sets}, \begin{math}O_1\;and\;O_2\;such\;that\;x_1\in O_1;\;x_2 \in O_2;\;and\;O_1 \cap O_2 \neq \varnothing\;where\; \varnothing\end{math}\emph{ is the null set}.
\\\\
To provide an example of a non-Hausdorff topology (that might also make understanding Hausdorff topologies easier), consider the following example that happens to be one possible manner in which a branching spacetime might be modeled.  Define a set \begin{math}\mathcal{E}_4\end{math} consisting of all the events of ordinary (3+1)-dimensional Minkowski space, \begin{math}M^4\end{math}.  Remove the set \emph{F} containing the spacetime event 0, and all subsequent events both inside and on the future light-cone with its vertex at 0.  Replace \emph{F} by two copies \begin{math}F_1\end{math} and \begin{math}F_2\end{math}.  The basis for such a topology on \begin{math}\mathcal{E}_4\end{math} is thus:
\noindent\begin{itemize}
  \item \emph{Any open set in} \begin{math}\left[M^4 - F\right] \cup F_1\end{math} \emph{is an open set in} \begin{math}\mathcal{E}_4\end{math}.
  \item \emph{Any open set in} \begin{math}\left[M^4 - F\right] \cup F_2\end{math} \emph{is an open set in} \begin{math}\mathcal{E}_4\end{math}.
\end{itemize}
\noindent It should be fairly clear why this topology fails to be Hausdorff, but Figure 3 \begin{figure}
\begin{center}
\includegraphics{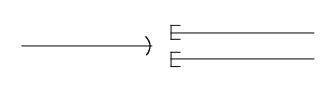}
\caption{A simple example of a non-Hausdorff topology represented by line splitting.  Adapted from \cite{visser}.}
\label{}
\end{center}
\end{figure}offers a visual description.  This is the essence of line-splitting and a generalized version of such a branched spacetime can be found in \cite{visser}.  This is also the underlying topology of the many-worlds interpretation of quantum mechanics. How this type of topology may be utilized will be discussed in the next section.  As such we are now in a position to introduce a protocol the utilizes CTC qubits to share information under the \emph{weak} consistency condition introduced above.

\section{OPERATIONS ON CTC QUBITS}
Bacon took a geometric approach to the discussion of quantum operations on CTCs by utilizing the Bloch sphere.  He also treated the CTC qubit much like Nielsen and Chuang treat the environment (see Ref. \cite{nc}, Ch. 8).  We take an algebraic approach and we simply treat the CTC qubit as a resource that can be utilized and manipulated (as long as weak consistency is maintained).  For our purposes here we will assume that the state of the CTC qubit is known.  What we will demonstrate is the feasibility of coupling an unknown state of some qubit to this CTC qubit in order to preserve and transmit that state for later measurement.

\subsection{Communication protocol: wavefunction analysis}
Suppose we have a qubit in some unknown state, \begin{math}|s\rangle=a|0\rangle+b|1\rangle\end{math}.  In fact, it is even possible that the basis is unknown here.  For example, this might represent a state in the \begin{math}\left(|0\rangle,|1\rangle\right)\end{math} basis or the \begin{math}\left(1/\sqrt{2}\left(|0\rangle+|1\rangle\right),1/\sqrt{2}\left(|0\rangle-|1\rangle\right)\right)\end{math} basis.  Suppose also that we have a CTC qubit initially in the known state \begin{math}|ctc\rangle=|0\rangle\end{math}.The state of the two qubits in the joint Hilbert space is \begin{equation} |s\rangle\otimes|ctc\rangle=a|00\rangle+b|10\rangle \end{equation}where we are in the basis such that \begin{math}|\alpha\beta\rangle\end{math} where \begin{math}|\alpha\rangle\end{math} represents the state of the chronology-respecting qubit (hereafter referred to as the chronology-respecting qubit) and \begin{math}|\beta\rangle\end{math} represents the state of the CTC qubit.  For the sake of description, we will have two people carrying out actions on the system at different times and locations whom we will call Alice and Bob.

Alice begins by passing the two qubits through a swap gate given by the unitary transformation \begin{math}\mathbf{U}=|00\rangle\langle00|+|01\rangle\langle10|+|10\rangle\langle01|+|11\rangle\langle11|\end{math}.  The joint state of the system after this action is \begin{equation}|j\rangle=U\left(|s\rangle\otimes|ctc\rangle\right)=a|00\rangle+b|01\rangle.\end{equation}  If we compare this to equation (5) we see that the swap gate has apparently accomplished this task.  Just to be sure, Alice will perform a measurement on the chronology-respecting qubit (\emph{after} the unitary transformation) in the \begin{math}\left(|0\rangle,|1\rangle\right)\end{math} basis.  We thus may rewrite equation (5) in terms of these measurements as \begin{equation}|j\rangle=\left(|0\rangle\otimes|v_1\rangle\right)+\left(|1\rangle\otimes|v_2\rangle\right)\end{equation} where \begin{math}|v_1\rangle\end{math} and \begin{math}|v_2\rangle\end{math} are \emph{unnormalized} vectors associated with the second (in this case the CTC) qubit \cite{loepp}.  These unnormalized vectors may be written in the form \begin{math}|v\rangle=\alpha|0\rangle+\beta|1\rangle\end{math}.  In setting equations (6) and (7) equal to one another, we find that \begin{math}|v_2\rangle\end{math} does not exist and \begin{math}|v_1\rangle=a|0\rangle+b|1\rangle\end{math}.  This means that Alice's measurement in the \begin{math}\left(|0\rangle,|1\rangle\right)\end{math} basis will put the chronology-respecting qubit into the state \begin{math}|0\rangle\end{math} and the CTC qubit will then be in the state \begin{math}a|0\rangle+b|1\rangle\end{math}, exactly what we expect from the swap gate.  Alice then communicates the result of her measurement to Bob.

At some later time and location (bounded by the knowledge of the CTC's size), Bob prepares a third qubit in the state \begin{math}|s^\prime\rangle=|0\rangle\end{math} based on Alice's reported measurement and then passes both the CTC qubit and his newly prepared chronology-respecting qubit through another swap gate.  This gives the joint state \begin{equation}|j^\prime\rangle=U\left(|s^\prime\rangle\otimes|ctc^\prime\rangle\right)=a|00\rangle+b|01\rangle.\end{equation}  If Bob then performs a measurement in the \begin{math}\left(|0\rangle,|1\rangle\right)\end{math} basis on the chronology-respecting qubit we may again rewrite this in a similar manner to equation (7).  We find that \begin{math}|v_1\rangle=a|0\rangle\end{math} and \begin{math}|v_2\rangle=b|0\rangle\end{math} meaning that, regardless of the values of \emph{a} or \emph{b}, the CTC qubit ends up in the state it was originally in before returning to Alice, thus preserving consistency.  Essentially, \emph{a} and \emph{b} simply determine the probabilities for the outcomes of this measurement.  This protocol is described in Figure 4. \begin{figure}
\begin{center}
\includegraphics{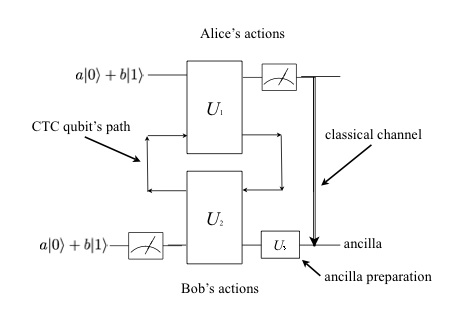}
\caption{This graphically depicts the protocol we describe.  Note that time flows from left to right for Alice and from right to left for Bob simply for the sake of preserving space.  For the CTC qubit and the classical channel, the arrows indicate the direction of time.}
\label{}
\end{center}
\end{figure}

\subsection{Communication protocol: density operator analysis}
Quantum operations are most frequently described in terms of density operators.  The protocol just described swaps the states of the two qubits before Alice performs a projective measurement on the chronology-respecting qubit in order to verify that it has indeed taken on the state of the CTC qubit.  In terms of density operators the new joint state may be represented as \begin{equation}\left(\rho_{out}\otimes\rho_{\textrm{{\tiny CTC}}}\right)=\frac{P_mU\left(\rho_{in}\otimes\rho_{\textrm{{\tiny CTC}}}\right)U^{\dag}P_m}{\textrm{Tr}\left(P_mU\left(\rho_{in}\otimes\rho_{\textrm{{\tiny CTC}}}\right)U^{\dag}P_m\right)}\end{equation} where \begin{math}P_m=|0\rangle\langle0|+|1\rangle\langle1|\end{math}.  The state of the chronology-respecting qubit is then obtained by tracing out the CTC in the denominator, \begin{equation} \rho_{out}=\frac{\textrm{Tr}_{\textrm{\tiny CTC}}\left(P_mU\left(\rho_{in}\otimes\rho_{\textrm{{\tiny CTC}}}\right)U^{\dag}P_m\right)}{\textrm{Tr}\left(P_mU\left(\rho_{in}\otimes\rho_{\textrm{{\tiny CTC}}}\right)U^{\dag}P_m\right)}.\end{equation}  Thus, if the CTC qubit was initially in the state \begin{math}\rho_{\textrm{\tiny CTC}}=|0\rangle\langle0|\end{math}, after the swap gate and projective measurement this same state should now be \begin{math}\rho_{out}\end{math}.  Alice, of course, reports this state to Bob who prepares a third qubit in this state, passing it through a swap gate with the CTC qubit, and performs the projective measurement on the chronology-respecting qubit resulting in the original state of the first chronology-respecting qubit being recovered as above.  In other words, Bob's actions should result in the state \begin{equation}\left(\rho_{out}^{\prime}\otimes\rho^{\prime}_{\textrm{{\tiny CTC}}}\right)=\frac{P_mU\left(\rho_{in}^{\prime}\otimes\rho^{\prime}_{\textrm{{\tiny CTC}}}\right)U^{\dag}P_m}{\textrm{Tr}\left(P_mU\left(\rho_{in}^{\prime}\otimes\rho^{prime}_{\textrm{{\tiny CTC}}}\right)U^{\dag}P_m\right)}.\end{equation}  Consistency is preserved and the process is identical to that described in the previous subsection as long as equations (9), (10), and (11) obey equation (4).

\subsection{CTC qubits as shared resources}
This motivates us to define the \emph{ctcbit} as a resource, being a qubit on a closed time-like curve shared by Alice and Bob.  By definition (via the Wheeler-DeWitt equation), CTCs are considered noiseless.  Note that Bob requires an ancilla, that is a generic qubit to be prepared in the state reported by Alice, in order to achieve the expected outcome.  Alice also communicates to Bob via a classical channel.  This is represented by a \emph{cbit}, a communication resource defined as being a classical bit sent over an equally classical channel.  In the quantum teleportation protocol, that accomplishes a similar end, Alice and Bob utilize an \emph{ebit}, a resource defined as being a maximally entangled state of a particular form for use in communication.  In that protocol, one \emph{ebit} and two \emph{cbits} produce one \emph{qubit}, where the latter is a communication resource equivalent to a physical qubit being sent over a noiseless channel.  The relationship between these resources in that particular protocol is \begin{equation}1\, ebit + 2\, cbits \ge 1\, qubit.\end{equation}

In the protocol we describe here, instead of using \emph{ebits} and \emph{cbits} to produce a \emph{qubit}, we use a \emph{ctcbit}, a \emph{cbit}, and an ancilla.  The relationship between these resources is then \begin{equation}1\, ctcbit + 1\, cbit + 1\,\textrm{ancilla} \ge 1\, qubit.\end{equation}  Note that Ref. \cite{vanenk} demonstrates that \begin{math}1\,\emph{qubit}\ge1\,\emph{ebit}\end{math}.  This, along with equation (13), implies that \begin{equation}1\,\emph{ctcbit}\ge1\,\emph{qubit}\ge1\,\emph{ebit}.\end{equation}  Note that, because the CTC and its associated qubit do not temporally evolve (outside of our forced unitary evolution), a CTC is truly a noiseless channel and well-suited for \emph{storing} information.  In other words, Alice might encode the information from her qubit in the CTC qubit as a way to protect it from degradation, etc.

\subsection{Utilization of non-Hausdorff topology}
We can now apply this simple idea to our basic communication protocol.  When BobÕs measurement is made, we implement a simple splitting of the CTC, and then perform the reverse (a Òline mergerÓ) in order to keep the histories consistent.  The latter can be visualized as Figure 3 running backward.  Mathematically let us define an open set \begin{math}\mathcal{E}_4\end{math} consisting of all possible states, any qubit can take both on a given Hilbert space, \begin{math}\mathcal{H}\end{math}, and on a simplified (3+1)-dimensional Riemann space, \begin{math}\mathbf{R}^4\end{math}, where \emph{m} = 4 + \emph{n} dimensions and \emph{n} is the dimension of the Hilbert space.  This means that any self-consistent single instance of an input state being successfully exchanged is represented by \begin{math}\mathcal{E}_m^\prime\end{math} where \begin{math}\mathcal{E}_m^\prime\subseteq\mathcal{E}_m\end{math} such that any two \begin{math}\mathcal{E}_m^\prime\end{math} do not intersect.  We consider the spatial location, or point, \emph{P}, of a \emph{ctcbit} to be part of its state.  Note that this works regardless of whether the Hilbert space is considered to be physically real or simply a mathematical construction.  We include the Hilbert space primarily for completeness of description.

So, for example, remove the point, \emph{P}, consisting of a spacetime point where the \emph{ctcbit} exits the swap gate, \emph{U}.  We will call this event 0.  Also remove the point, \emph{Q}, consisting of a spacetime point where the \emph{ctcbit enters U}.  We will call this event 1.  Additionally, remove all events both inside and on the future light-cone of \emph{P} as well as inside and on the past light-cone of \emph{Q}.  Note that these are the same since it is a CTC!  Replace the set of all events between and including \emph{P} and \emph{Q} with \emph{i} copies such that any set in \begin{math}\mathcal{E}_m^\prime=\left[\mathbf{R}^4 - \left\{P,Q\right\}\right]\cup\left\{P_i,Q_i\right\}\end{math} is a set in \begin{math}\mathcal{E}_m\end{math}.  This is diagrammatically represented in Figure 5. \begin{figure}
\begin{center}
\includegraphics{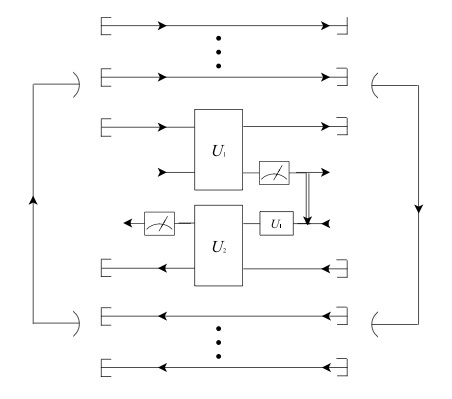}
\caption{This demonstrates the non-Hausdorff topology when applied to the CTC.  The arrows give the direction time flows at any given point.  Note that Figure 2 is essentially embedded in the center.  The vertical elipsis indicates that there may indeed be an infinite number of branches.  The idea is that there is only \emph{one} branch that is accessible to Alice and Bob.}
\end{center}
\end{figure}

\section{CONCLUSION}
There a re two major fundamental points we have introduced here.  The first is that we have noted that the traditional interpretation of equation (3) is making the CTC qubit dependent upon the chronology-respecting qubit to which it is coupled is not necessarily the only interpretation of that equation and, in fact, a weaker and less paradoxical interpretation does exist.  Additionally we have shown that by employing a weaker consistency requirement than Bacon's and Deutsch's, in the form of equation (4), it is possible to utilize qubits on closed time-like curves as a communication resource, even allowing for a changing state.  Consistency, we have shown, may be logically maintained as long as the state remains the same at any given point on a given CTC.  This also assumes that time on CTCs has a chiral nature and that it only flows parallel to linear 'background' time (i.e. causal time) when the two are coupled via a unitary transformation or some such action.  To further ensure consistency and prevent limitless use of the resource we employ a non-Hausdorff topology, only allowing Alice and Bob access to the CTC qubit once.  This is consistent with Visser's suggestion that such a topology would be helpful in preserving causality in the presence of CTCs \cite{visser}.

We also analyzed the CTC qubit as a shared information resource and demonstrated its relation to other information resources.  One might legitimately ask what advantage, if any, a \emph{ctcbit} might have over, for instance, an \emph{ebit}.  As noted in Refs. \cite{fuentes05,alsingmilburn,haydenpreskill}, gravitational fields can degrade entanglement.  Since CTCs likely include a wormhole throat somewhere on the curve, and such a throat is most likely caused by a black hole, our protocol might provide a useful alternative to quantum teleportation in the presence of such very strong gravitational fields since no \emph{ebit} is required.  This still presents the problem of gravity potentially altering the \emph{ctcbit} itself but hints that there might be a limit where this protocol might work in a slightly weaker gravitational field that would otherwise erode entanglement.  This is clearly an area of work that needs to be addressed in the future.

We also restricted ourselves to the use of swap gates.  Bacon has demonstrated consistency for CTC qubits involved in other unitary transformations including a controlled-phase gate followed by an exchange of two qubits as well as a controlled-rotation gate, though the coupling between the CTC qubit and CRQ qubit is much stronger than in the simple swap gate and thus the type of information that may be exchanged is more heavily constrained.  Nonetheless, it might be interesting to see what sorts of combinations of quantum operations could be performed with CTC qubits under the conditions we have set forth.

Finally, we noted that, due to the essentially noiseless character of CTCs, this may prove to be a useful way to store information for extended periods of time.  In other words the CTC acts a bit like a data storage device.  Conversely one might also hold that this allows the data to be protected in a way, i.e. from degradation or interference.  Of course, all of this is predicated not just on the existence of CTCs but the ability to manipulate them.  It might be possible to relax the requirement of a non-Hausdorff topology (or, perhaps, apply it only to the Hilbert space, whatever that might mean physically), but one would still be left with the need for a CTC.  Nonetheless, it presents an interesting theoretical laboratory in which we can study the relationship between quantum communication and gravity, and also suggests that perhaps an alternative to quantum teleportation exists that does not require an entangled pair.

\begin{acknowledgements}
We are greatly indebted to Barry Sanders and Dave Bacon for assistance on various drafts, to Patrick Hayden for both an enlightening discussion and suggestion for future work, and to Mark Wilde and Todd Brun for continuing encouragement and helpful comments.  We also acknowledge helpful discussions with Ivette Fuentes-Schuller, Ken Wharton, Chris Altman, and Lorenza Viola.
\end{acknowledgements}

\bibliographystyle{apsrev}
\bibliography{DurhamBibliography.bib}

\end{document}